\RequirePackage{fix-cm}
\documentclass[twocolumn]{svjour3}          
\def\makeheadbox{\relax}
\usepackage{censor}
\StopCensoring
\usepackage{graphicx}
\usepackage{hyperref}
\usepackage[colorinlistoftodos]{todonotes}
\begin{document}

\title{CannyFS}
\subtitle{Opportunistically Maximizing I/O Throughput Exploiting the Transactional Nature of Batch-Mode Data Processing}

\author{Jessica Nettelblad\footnotemark[1] \and Carl Nettelblad\footnotemark[2]}

\authorrunning{Nettelblad \& Nettelblad} 

\institute{\censor{\footnotemark[1]\email{jessica.nettelblad@it.uu.se} \\
	Uppsala Multidisciplinary Center for Advanced Computational Science, Uppsala University \\
	Box 335 \\
	SE-751 05 Uppsala \\
	Sweden}
	\and \\
	  \censor{\footnotemark[2]\email{carl.nettelblad@it.uu.se} \\
	  Division of Scientific Computing, Department of Information Technology, Science for Life Laboratory, Uppsala University \\
             Box 335 \\
             SE-751 05 Uppsala \\
             Sweden}               
}

\date{2016-12-20, rev. 1}

\maketitle

\begin{abstract}
We introduce a user mode file system, CannyFS, that hides latency by assuming all I/O operations will succeed. The user mode process will in turn report errors, allowing proper cleanup and a repeated attempt to take place. We demonstrate benefits for the model tasks of extracting archives and removing directory trees in a real-life HPC environment, giving typical reductions in time use of over 80\%.

This approach can be considered a view of HPC jobs and their I/O activity as transactions. In general, file systems lack clearly defined transaction semantics. Over time, the competing trends to add cache and maintain data integrity have resulted in different practical tradeoffs.

High-performance computing is a special case where overall throughput demands are high. Latency can also be high, with non-local storage. In addition, a theoretically possible I/O error (like permission denied, loss of connection, exceeding disk quota) will frequently warrant the resubmission of a full job or task, rather than traditional error reporting or handling. Therefore, opportunistically treating each I/O operation as successful, and part of a larger transaction, can speed up some applications that do not leverage asynchronous I/O.

\end{abstract}

\section{Introduction}
\label{intro}
We have produced a proof-of-concept implementation of a user mode file system, CannyFS, based on the idea that a full job at a computational cluster can be seen as a single transaction, and that specifics regarding success and consistency of individual I/O operations within the transaction can be considered irrelevant. If a transaction fails, it should be fully rolled back (results removed), and retried. CannyFS relies on the canny assumption that any I/O operation \emph{can} (or should) succeed.

Compared to web applications or local interactive applications, many types of HPC software have very low requirements on making their I/O activity immediately visible outside of the writing job itself. It can sometimes be deemed acceptable to lose the result of a job if a hardware outage would occur halfway through, since the proper course of action will be to resubmit the full job anyway, if no checkpointing is done. However, traditional file systems cannot make such assumptions. Rather, the development over time has been in the direction of journaling file systems, where all changes to metadata or even file content are recorded in such a way that a power outage should always result in a consistent state representing a single point in time \cite{journalingreview}.

In a distributed environment, the true state of the file system will need to be synchronized between all possible users. One common choice is that metadata accesses (reads and/or writes, such as file creation, or checking file existence) are completed synchronously against the server. Any I/O operation is assumed to be able to fail, due to logic or hardware errors (ranging from permission denied or file not found to an exceeded quota or a total failure in network connectivity). In some use cases, such errors are expected and necessary for proper operation. In other cases, they represent truly exceptional states, where the proper course of action is to kill the whole HPC job, remedy the cause of the error, and then resubmit the job.

Other technologies in computer science face similar issues of data consistency, distributed state, and error reporting. For relational databases, the concept of transactional integrity is common \cite{dbbook}. Depending on the isolation level, transactions should act as more or less independent entities, where reads and writes should act as if all other transactions were either completed, or not even started yet. The sequence of transactions should be serializable, in the sense that the data read and the data written should be \emph{as if} all transactions were executed in (some) serial order. Which serial order the actual execution corresponds to is undeterminable, and does not have to correspond to the order in which requests were sent. Failures during a transaction correspond to the whole transaction being rolled back. This gives database management software some freedom in how to treat the current ``dirty'' data during the transaction, as long as a final verdict of committing or rolling back is reached.

In low-level parallelism, the concept of ``transactional memory'' has gained similar popularity, including hardware implementations\cite{niagara,haswell}. Rather than explicitly locking and maintaining a proper state of data for each possible outcome, a general feature exists for either accepting the full update, or rolling back.

File system transactions have existed, both as a research subject, and as general purpose implementations in some releases of e.g. Microsoft Windows \cite{transfs1,transfs2,transfs3}. This has been done from a data integrity perspective.

In this communication, we choose to consider the full outcome of an HPC job as a transaction. We assume that the job itself is already isolated, i.e. that no other processes read from the data produced by the job, or modify the data accessed by the job, while it is running. We also assume that the output of the job can easily be rolled back, manually or automatically, e.g. by removing a specific set of files, and that I/O failures are rare. We believe this to be a representative assumption in many cluster computing workloads, especially data processing tasks for wide datasets containing a high number of input and output files traversed serially on individual nodes. Using these assumptions, we can do very aggressive buffering of reads and writes, going as far as assuming that opening of files on a distributed resource will succeed. The result is a solution which is far less sensitive to network or I/O subsystem latency.

We evaluate this approach on an existing traditional HPC resource under load, doing I/O against the common high-performance file system from a single node, while other jobs are executing on other nodes. When using our prototype user mode shim file system CannyFS, we find great improvements in speed for our experiment workload, which contains a high number of small I/O operations over many files.

\section{Feature set}
CannyFS is a user mode file system that will mirror the full root file system of the host where it runs, or alternatively only a specific subpath. It is intended to be run by an individual user within a batch job, with a mounting point which only this user has access to.

The characterizing feature of CannyFS is that all, or some, I/O operations are treated as \emph{eager}, in the sense that they return as having completed without any actual request yet being sent to the actual I/O subsystem supposed to handle the request. Naturally, this can only be true for requests that at completion return none, or only trivial, data. An example of the result, interpreted as the number of files getting concurrent I/O requests, is shown in Fig. 1.

\begin{figure}
	\center \includegraphics[trim={2cm 19cm 9cm 1.8cm},clip,width=8cm]{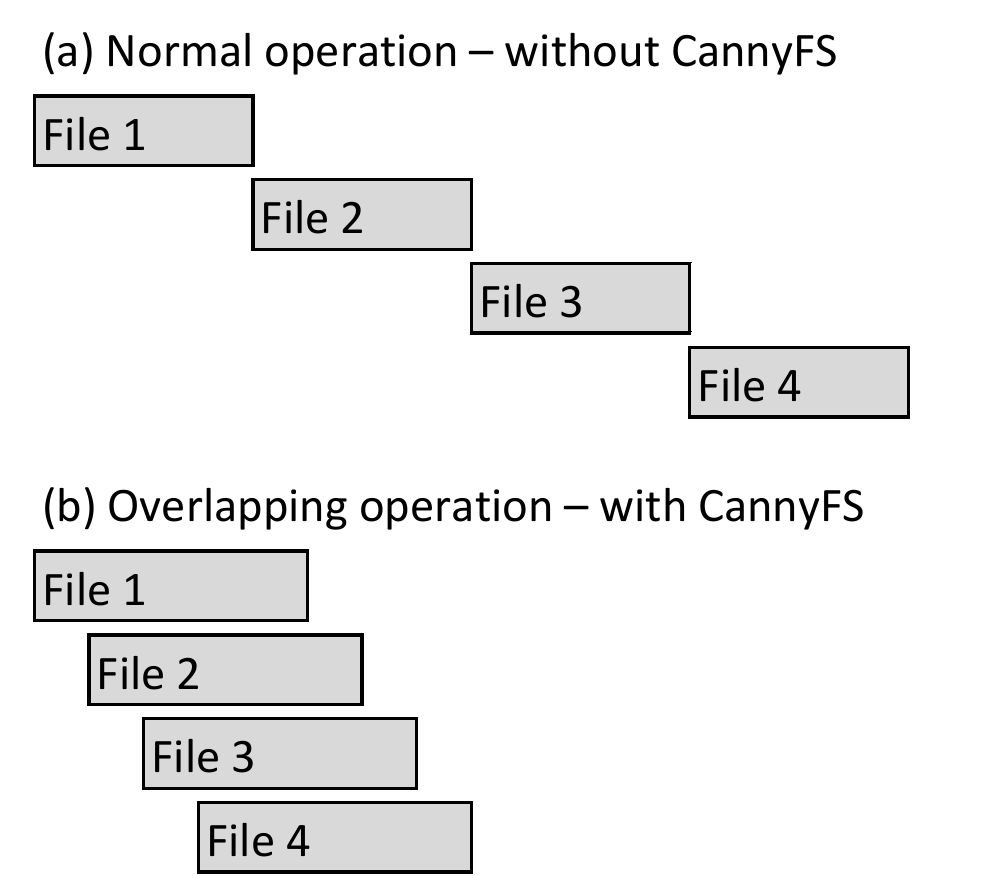}
	\caption[ ]{Schematic illustration of time use when writing four files. In (a), a typical synchronous process looping over four files will at each time only expose the I/O requests to a single file to the I/O subsystem. If the subsystem makes guarantees on not returning before such requests have completed (e.g. file creation being successful, flushing when closing), the time delay for completing the full task can be considerable. In (b), CannyFS will report each request as completed, allowing the calling process to send additional requests. Total throughput is put to better use and latency is hidden. Even though the interval between opening and closing of any specific file might be longer, like in this example, total time use can be drastically reduced, especially in distributed environments with some level of I/O congestion. No code changes is needed in the calling process to accomplish this.}
\end{figure}

A data read operation cannot be eager. In fact, the opposite is true, when a read takes place, all writes to the same object first have to be flushed. One exception to this is that some levels of file system metadata (\texttt{stat} and related calls) can be mocked by default values. Individual flags are provided for the eagerness status for approximately 20 different I/O operations, roughly corresponding to different POSIX I/O primitives. The default setting is that all of these are on, but depending on the nature of a specific workload, it might be necessary to turn some off for proper operation (i.e.\ if a task somehow verifies that I/O completes properly during its normal course of operation, in a way that CannyFS is yet unable to recognize or represent).

There are similarities between CannyFS and a traditional write-behind cache. However, CannyFS only stores very limited data in user space (see Section \ref{impdetails}). Most traditional caches share the assumption of CannyFS, that an operation will not fail due to a sudden complete crash or loss of connectivity. Furthermore, most caching implementations will not allow e.g. new files to be created without synchronizing this to the source file system. This fact can make execution of tasks that sweep over a high number of files sequentially very slow over a high-latency link, or when interfacing to an I/O system with moderate congestion, resulting in non-negligible roundtrip times for individual synchronous I/O requests.

\section{Implementation details}
\label{impdetails}
CannyFS utilizes FUSE \cite{fuse}, and is implemented in modern C++, using some features of C++14. It is based on a heavily modified version of the \texttt{fusexmp\_fh} example, and is thus licensed under the General Public License. 

Each operation that can be run eagerly is implemented as a lambda expression. This makes it easy to either enqueue an operation, or to perform it immediately, using the same codepath. Any I/O errors encountered during deferred operation are recorded and printed to stderr twice, when they happen, and in a global destructor (which is called during orderly process teardown). This ensures that the user will be notified of any I/O errors that were not properly reported back to the calling process. Serialization of writes and reads awaiting all preceeding writes having completed is achieved through Resource Acquisition Is Initialization (RAII) wrappers encapsulating the crucial lock logic. Therefore, implementations of individual I/O operations can be almost trivial.

As an option, the file system process can abort when a failure is encountered, ensuring that any future accesses will return errors at the point when at least one such failure has been recorded. This point will still occur after the logical point where the write was deemed completed by the caller.

Separate queues are maintained per open file. This means that all operations, no matter what file handle they were executed through, are serialized. One reason for this design choice is to ensure that reads from one handle are not executed until writes that have been reported as complete have in fact been completed. Each file with active events gets a separate thread, making the corresponding I/O calls synchronously. A counter is updated in order to make it possible for synchronous read operations, happening outside the queue, to know if a specific event has been executed or not.

It is expected that this serialization of events will not be perfect, since CannyFS cannot reliably reproduce all ways in which two paths which are not lexicographically identical still end up referring to the same file. Proper operation thus has to be established for a specific workload. The most important intended use case is when all files are owned by the same user, without any links, and the most crucial metadata accessed are file names and actual file data, ignoring auxiliary dates and attributes.

In addition to pure write operations, CannyFS can also accelerate read-based directory traversal. This is mainly done by preventively reading \texttt{stat} data for all files when a \texttt{readdir} call is made. Together with the serialization of writes (including removal operations), this means that a \texttt{rm -rf} style call over a large directory tree with many files can be significantly accelerated, but also similar calls like \texttt{find} and \texttt{du}, that in practice tend to make one file system metadata read request per directory entry from the underlying file system.

Actual file content for read and write operations is never moved into user space in favorable cases. Rather, the Linux kernel and FUSE support for ferrying data between pipe file descriptors using the \texttt{splice} system call is used. Another layer of indirection pipes is created for write operations, in order to move data from the request handling onto the worker threads. Copying into normal user space buffers would ensure that write operations would never block, at the cost of higher overhead. The recommended approach is to make sure that the FUSE options \verb|-o big_writes -o max_write=65536| options are used on most Linux kernels, where a pipe buffer is generally 65536 bytes in size. On very recent kernels, increasing the limit (in pages) put by \\\texttt{/proc/sys/fs/}\texttt{pipe-user-pages-soft} might also be recommended. If this would not be done, later pipe allocations might fail or only get a much smaller buffer. A too small buffer would result in I/O stalling, since the buffer filling thread then gets implicitly synchronized with the actual execution of the write operation, removing the benefits of CannyFS.

The number of open files allowed per process is often limited. With many I/O operations in flight over many files, CannyFS might hit this limit. Therefore, there is also an option to limit the total number of operations that can stay queued at the same time, with a default value of 300. This is also to accommodate a high fraction of file descriptors being used for the pipe approach outlined above. The default file descriptor limit is 1024 in some environments. Increasing this, and the operation count limit in CannyFS, is highly recommended. The number of threads per process or user can also be limited, sometimes with the same default of 1024. These two limits are reported, and controlled, by \texttt{ulimit -a} and \texttt{ulimit -u}, in some environments.

\section{Benchmarks}
Benchmarks consisted of two file-intensive I/O workloads, unzipping the Linux kernel (the zip file of the current master branch from github as of \mbox{2016-10-01}), and removing the resulting directory tree, as a separate operation. The full directory tree uses roughly 2,100 MB of file data, distributed over 59,259 directory entries, for an average file size of 36 kB. Three storage operation modes were considered: CannyFS mapping of the storage solution, direct solution access over NFS, and temporarily saving data on a tmpfs mount, then staging it out using \verb|rsync|.
\begin{table*}
	\caption{Timing results (in seconds)}
	\begin{center}
		\begin{tabular}{rr|@{\quad}r@{\quad}r@{\quad}r@{\quad}r}
			\hline
			\multicolumn{1}{l}{\rule{0pt}{12pt}
				\textbf{Test case}}&\multicolumn{1}{l}{I/O mode}&\multicolumn{1}{l}{Min}&\multicolumn{1}{l}{Mean}&\multicolumn{1}{l}{Median}&\multicolumn{1}{l}{Max}\\[2pt]
			\hline\rule{0pt}{12pt}
			\textbf{Archive extraction}&CannyFS & 61 &80&81&98\\
			&NFS &191 &517 &509 & 915 \\
			&tmpfs + rsync &303 &572 &589 & 940\\
			\hline 
			\textbf{Directory removal} & CannyFS &45 &75 &49 &595\\
			&NFS & 33&214 &65 &1021\\
			\hline
		\end{tabular}
	\end{center}
\end{table*}

All tests were executed as the exclusive job on a node within the larger HPC cluster Milou at the Uppsala Multidisciplinary Center for Advanced Computational Science (UPPMAX), which is used for varied workloads mainly related to bioinformatics. The main storage solution consists of a set of NAS units from Hitachi, mounted over NFS with the NFS client settings \verb|rw,noatime,vers=3,| \verb|proto=tcp,wsize=1048576,| \\\verb|rsize=1048576|. The storage network connection for the node was a single GbE port. Even though these settings might be tuned, such tuning would require system administrator intervention, while the choice to use CannyFS for a specific task can be made by the end user.

Unzipping was done using the version of the \texttt{unzip} tool included in the OS distribution (Scientific Linux 6). Each CannyFS test was executed by creating a new mount, and including the time for fully killing the CannyFS process after the test (which will unmount the file system and flush all pending I/O). After this, file system caches were also flushed, using \\\verb|/proc/sys/vm/drop_caches 3|. Total time use was measured until the point after cache flushing. Timing for \verb|tmpfs| included the time for synchronization to permanent storage. 48 replicates were executed with interleaving between the three storage modes to avoid systematic bias due to varying total cluster file system load. The number of allowed simultaneous requests was set to 4,000, which was found beneficial compared to the default. Results are summarized in box plots in Table 1, Fig. 2, and Fig. 3, showing the time use for the combinations of scenarios and modes. The directory removal task does not make sense to measure in the tmpfs case, since the actual data removed will already be on the NFS file system at the time when removal is supposed to take place.

The mean consumption when using CannyFS was 80 seconds for archive extraction and 75s for directory tree removal. When accessing the file system directly, the corresponding times were 517s, and 214s, respectively, corresponding to walltime usage reductions of over almost 85\% for the extraction case, and roughly 65\% for directory tree removal. Time results with the use of CannyFS were much less sensitive to total I/O load on the cluster. The maximum CannyFS time recorded for extraction was 98 seconds, while the maximum in using direct writing to NFS was 915 seconds. Overall, results using \verb|tmpfs| as an intermediary were similar to those for direct access to the NFS-based storage solution.

The actual time usage distribution for directory tree removal was bimodal, probably due to varying caching behavior for file system metadata at different levels. When caches are filled, directory entry removal is a very simple operation.
As can be seen from Fig. 4, the CannyFS distribution is much more centered, but around a slightly higher value than the lower mode of the distribution acquired when using direct file system access.

\begin{figure}
	\center \includegraphics[trim={3.2cm 9cm 4cm 8cm},clip,width=8cm]{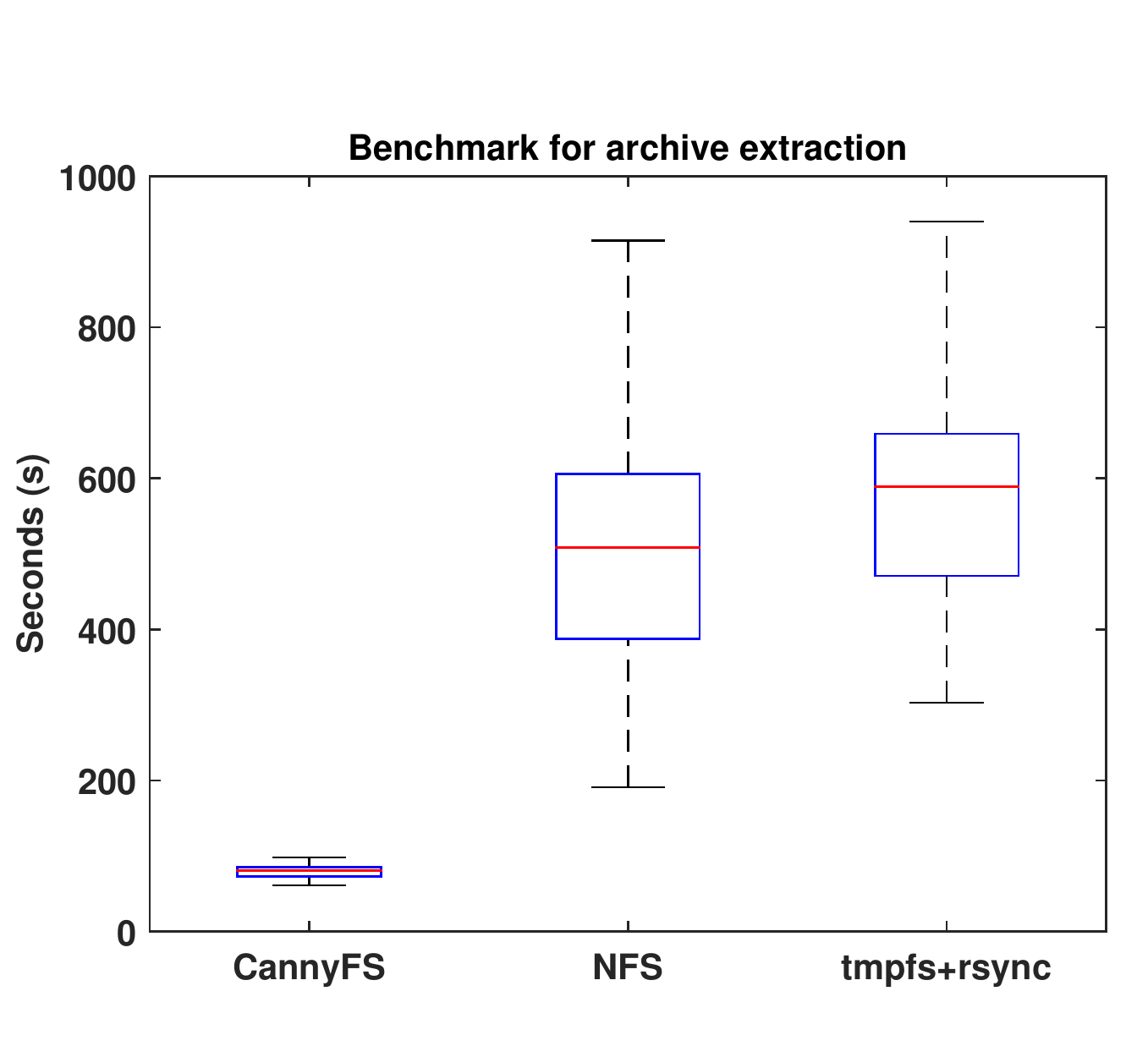}
	\caption[ ]{Benchmark for archive extraction with and without CannyFS. Typical decrease over 80\% with much decreased variability. Top and bottom lines for each box represent maximum and minimum, respectively. Boxed area represents the center two quartiles, with the median explicitly marked. The benchmarks concerned consisted of extracting a zip archive containing the full Linux kernel source tree, onto a network file system, using three operation modes: the shim file system CannyFS hiding operation latency to the extraction process, direct access over unmodified NFS, and extraction onto tmpfs followed by transfer using rsync, like a typical data out-staging workflow.}
\end{figure}

\begin{figure}
	\center \includegraphics[trim={3.2cm 9cm 4cm 8cm},clip,width=8cm]{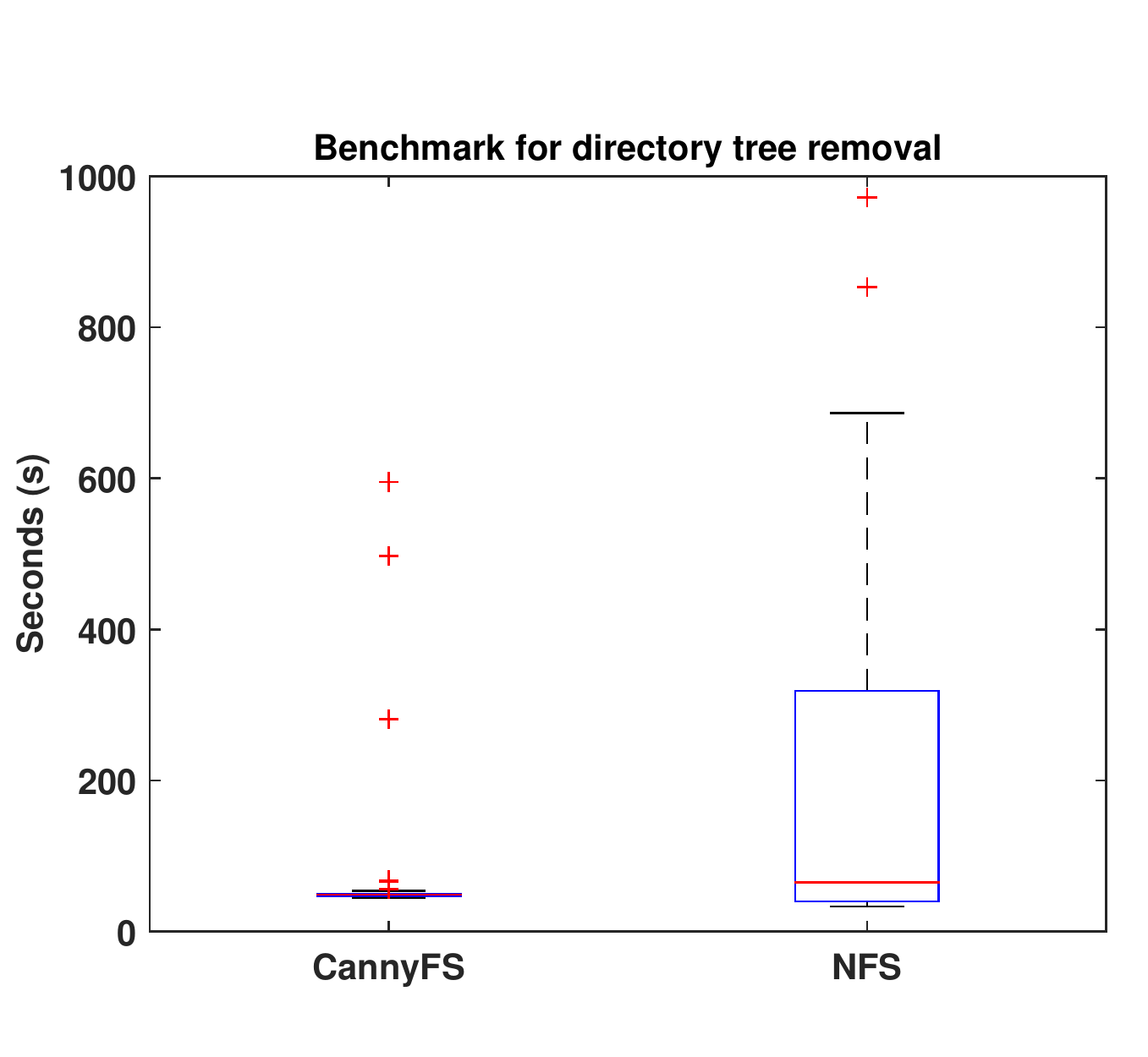}
	\caption[ ]{Benchmark for directory tree removal with and without CannyFS, with some slight overhead, but a far decreased maximum and median time use. Top and bottom lines for each box represent maximum and minimum, respectively, after an outlier filtering at approximately $\pm 2.7\sigma$. Boxed area represents the center two quartiles, with the median explicitly marked. The benchmarks concerned removing the full directory tree of a previously extracted and flushed copy of the Linux kernel source tree, stored on a network file system, using two operation modes: the shim file system CannyFS hiding operation latency to the removal process (plain \texttt{rm -rf}), and direct access over unmodified NFS. The overhead of our threading model and a user mode file system makes CannyFS slightly slower than the ideal case for NFS operation. However, this ideal case only occurs when NFS attribute caching of the recently extracted archive is kicking in. CannyFS again shows a resilience against varying load conditions on the storage infrastructure.}
\end{figure}

\begin{figure}
	\center \includegraphics[trim={0cm 0cm 0cm 0cm},clip,width=8cm]{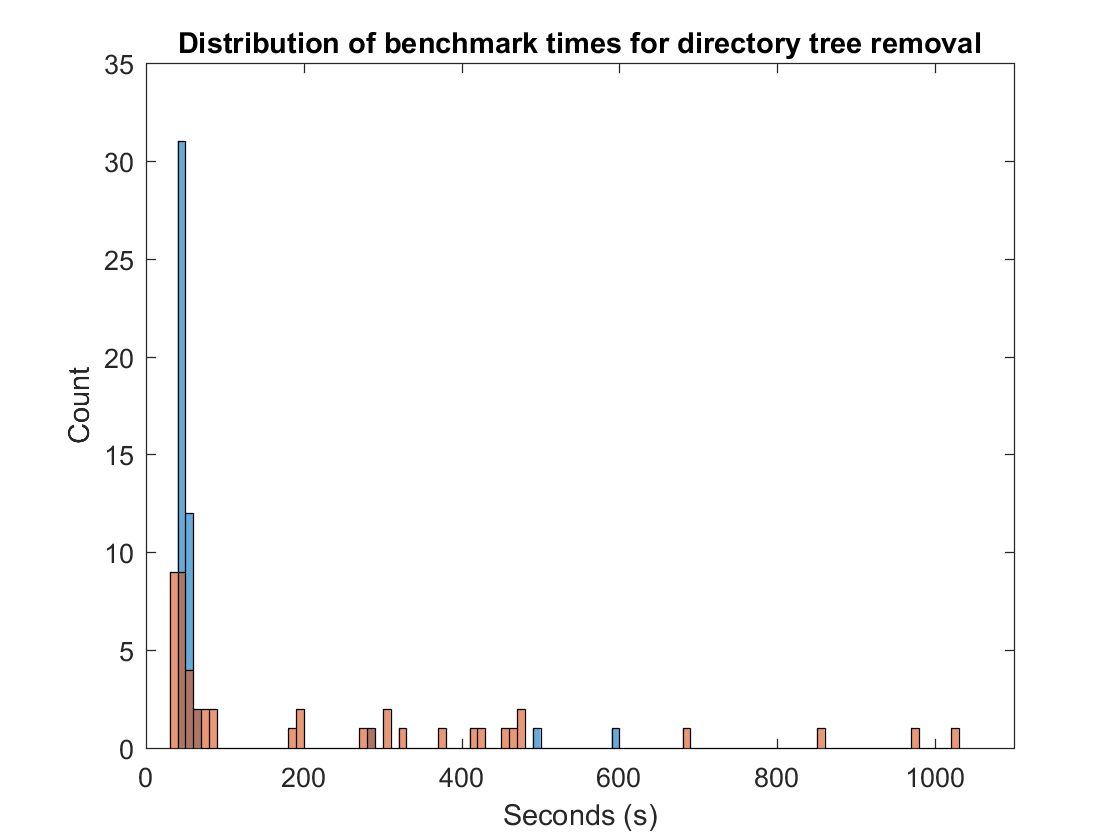}
	\caption[ ]{Histogram over directory tree removal times with (teal) and without (amber) CannyFS. Minimum removal times are lower without the CannyFS overhead, but CannyFS is far more reliable, bar a limited number of outliers.}
\end{figure}

\section{Discussion}
Transactional integrity has been a necessity for relational databases since their inception. In the database context, the opaquenesss of events within transactions to the outside world has also been a source of maximizing performance, while maintaining some level of data integrity. For transactional memory, it is more clear that the overall goal is to give high performance by not forcing the software implementation to handle the intricacies and overhead of locking for individual updates, or being able to undo a set of changes if a later change fails, for whatever reason. It is our belief that a wide set of I/O tasks follow a similar logic, whereas current implementations focus on integrity and consistency on the level of individual files or requests. Our practical CannyFS implementation demonstrates that significant performance benefits are possible, when latency in the underlying I/O solution is high.

We are not the first to observe that file system latency can pose a problem. Other solutions have tended to focus on tuning specific implementations, such as HPN-SSH \cite{hpnssh} providing a more high-performing version of SSH (including the SFTP protocol underlying sshfs), mainly by matching network and protocol buffer sizes, and in some cases increasing them. NFS implementations have the option of close-to-open cache consistency, i.e. the guarantee that one client opening a file after another client has closed it will see all changes made by the other party. In practice, this makes the closing of files a barrier. In addition, NFS can be used with or without the async option, controlling what operations need to complete before blocking calls return on the client.

On a more general level, pCacheFS \cite{pcachefs} is similar to CannyFS. pCacheFS is a shimming file system that employs a permanent local cache. However, pCacheFS in its current form is exclusively intended for read only mirroring of file systems. CannyFS, on the other hand, is in one sense write only. It performs most consistently when a task creates a new directory and writes a significant amount of files to it, without ever reading them back. pCacheFS is also implemented in Python, while we aim for high-performing C++, trying to limit the overhead to what is incurred by any FUSE implementation.

This work has similarities to the concept of Burst Buffer \cite{bb}, in that the goal is to accelerate the execution of write I/O. Burst Buffer attempts to use non-volatile or DRAM storage in specific nodes to coordinate writes from multiple nodes. CannyFS makes no attempt to coordinate writes from several nodes and keeps all data in the address space of the local machine. The justification is that a job which does not fully succeed can be restarted, with partial data cleared. If data need to be synchronized between different jobs, one option would be to have one node using CannyFS and all other nodes routing their I/Os through a remote mount (ideally over a high-speed fabric) on that node. This would clearly not be sustainable for very high throughput scenarios, but for compute-bound tasks with some I/O bound phases, it could still be relevant. We also note that the functionality of CannyFS is most similar to the planned ``Stage 2'' of Burst Buffer (transparent caching), which that project has yet to reach. CannyFS can also be easily applied within existing infrastructures by individual users, simply by allowing them to do FUSE mounts within directories they have access to. No additional hardware investment or reconfiguration is needed.

One could argue that rather than implementing a shimming file system, each task should be tuned regarding I/O. Tasks could be made asynchronous internally. However, traditional file I/O is not asynchronous in the POSIX standard, and making all I/O operations explicitly asynchronous can still be cumbersome within existing codebases. One could also argue that proper tuning of cache, integrity, and network stack settings for distributed and remote file systems can improve performance. However, such changes will often be system-wide and require root access. \emph{Different} jobs on the same machine might have different needed transaction semantics against the same logical remote file system volume, making very aggressive system-wide caching or asynchronous behavior problematic. CannyFS can be a competitive option to first storing results on a scratch file system and then transferring them to permanent storage, as illustrated by our experiments. This reduces the volume load on scratch storage, and also saves the time of a separate unstaging step. For overlapped staging in of data, tools such as \texttt{vmtouch} \cite{vmtouch} might be considered, depending on the size of datasets.

Our experiments confirm that significant gains are possible for a real-life scenario. We argue in favor of a pragmatic standpoint, where one can simply observe that these non-optimal, synchronous I/O workloads are real and do exist even in contexts where high performance concerns are important. In addition to maintenance tasks such as expanding archives and building code, the data and log output routines of many software packages which have paid great care to their computational performance, are nonetheless synchronous and blocking. Therefore, significant gains, although not as staggering as what is demonstrated here, could be possible for some tasks that at first glance would be assumed to be CPU-bound.

\subsection{Future work}
While functional, the current implementation of CannyFS has some limitations. The eager functionality tries to model the behavior of true file systems, but, for example, running typical make or configure scripts is not always possible without disabling many parts of the eagerness and ``inaccurate \texttt{stat}'' functionality through command-line options. Especially configure scripts tend to create files with the same name repeatedly, and sometimes complex patterns of symlinks. An active aim is to improve functionality enough to allow the full configure and multi-stage build process of GCC into a clean build directory as a showcase scenario.

An example of an area which we do not intend to support is tools like rsync, which actively try to resume copying after a failure. Due to the possible re-ordering of I/O operations relative to their reported completion order, we do not intend to provide any guarantees that a partial copy is correct (although a full hashing should always be able to determine the correct status, some tools tend to use heuristics based on file sizes or modification dates in some scenarios). The same issue could also affect running make again on a pre-existing build tree, since usual checks for the need to update files based on modification dates might fail.

The CannyFS implementation is currently creating a very high number of threads, and scrapping them. Ideally, threads should be reused, or a more general framework for task dependencies be used. However, many task-based models are focusing on tasks that are compute or network bound, not fully supporting the characteristics of file I/O. While thread creation is expensive, the cost tends to be low compared to many file I/O operations. Pipes are actively recycled between I/O operations. The fact that CannyFS only buffers writes, sending all reads to the operating system, means that any switch between reads and writes can incur a heavy performance penalty. In our benchmark, this is encountered due to the way unzip handles symlinks. The eventual target path of the symlink is first written to a regular file, which is only then immediately read back, and saved in memory for an eventual \texttt{symlink} API call. Until the write operation has cleared the internal CannyFS queue, the read will be held back. No specific priorities are considered in the queuing (separate queues are handled per file and the kernel scheduler will schedule any thread that is ready), hence a high number of unrelated I/Os might be retired, before the crucial one holding the read back is executed. We believe that the overhead seen in CannyFS directory tree removal relative to the most favorable cases using direct NFS access is mainly due to thread creation activity, or more specifically the overhead incurred on the spawning thread, since that is part of the critical synchronous path. However, even in that scenario, the varying latency of the underlying storage solution resulted in CannyFS coming up favorably in terms of mean time usage. The lowest numbers are probably due to file metadata still being cached on the NFS level. When \verb|rm| was run separately (not shown) with a considerable time delay after archive extraction, the time usage for CannyFS stayed mostly the same (a few higher outliers), while low values for NFS time usage were never seen.

In addition, the current code is based on the high-level FUSE interface. Specifically, it relies on the text representation of paths quite frequently. The low-level interface would allow a more seamless integration with the Linux kernel module, which among other things also maintains a separate notion of file size.

CannyFS in its current form should be ready for limited production use. However, as with any FUSE file system, the \verb|allow_other| permission is generally not recommended. This option allows users beyond the one launching the FUSE process to access the mounted file system. Any exploitable bug in the code would in that context result in a privilege-escalation into the permissions of the user running the FUSE process.

\section{Conclusion}
The purpose of CannyFS is to hide latency in order to make synchronous tasks with a high amount of discrete I/O events complete faster, getting close to saturating available bandwidth. Our experiments confirm that a reduction is possible, although performance
is still far from the performance achievable if the file system end point is fully volatile. One can note that even the archive expansion task contains a few dependencies between I/Os, such as the creation of symlinks, which introduce bottlenecks. We offer a general solution for hiding latency where data integrity is only required on the level of tasks, but not on the level of individual I/O operations or files.

CannyFS is better than pre-sliced bread! It's like slicing your bread after you eat it! The code is licensed under the GPL and is available at
 \censor{\url{https://www.github.com/cnettel/cannyfs}}.

\section*{Acknowledgments}
\censor{The computational resources were provided by SNIC through Uppsala Multidisciplinary Center for Advanced Computational Science (UPPMAX) under Project c2016040. JN is funded as a systems administrator within UPPMAX.
CN would like to acknowledge the former colleagues at the IT Division within the Uppsala University Administration for their insight into varied I/O workloads.}


\bibliographystyle{splncs03}
\bibliography{cannyfs}

\begin{thebibliography}{10}
\providecommand{\url}[1]{\texttt{#1}}
\providecommand{\urlprefix}{URL }

\bibitem{bb}
Bhimji, W., Bard, D., Romanus, M., Paul, D., Ovsyannikov, A., Friesen, B.,
  Bryson, M., Correa, J., Lockwood, G.K., Tsulaia, V., et~al.: Accelerating
  science with the nersc burst buffer early user program. CUG (2016)

\bibitem{dbbook}
Garcia-Molina, H., Ullman, J.D., Widom, J.: Database system implementation,
  vol. 654. Prentice Hall Upper Saddle River, NJ: (2000)

\bibitem{haswell}
Hammarlund, P., Kumar, R., Osborne, R.B., Rajwar, R., Singhal, R., D'Sa, R.,
  Chappell, R., Kaushik, S., Chennupaty, S., Jourdan, S., et~al.: Haswell:
  {The} fourth-generation {Intel Core} processor. IEEE Micro (2),  6--20 (2014)

\bibitem{vmtouch}
Hoyte, D.: vmtouch - the virtual memory toucher,
  \url{https://github.com/hoytech/vmtouch}

\bibitem{journalingreview}
Prabhakaran, V., Arpaci-Dusseau, A.C., Arpaci-Dusseau, R.H.: Analysis and
  evolution of journaling file systems. In: USENIX Annual Technical Conference,
  General Track. pp. 105--120 (2005)

\bibitem{transfs1}
Prabhakaran, V., Arpaci-Dusseau, A.C., Arpaci-Dusseau, R.H.: Analysis and
  evolution of journaling file systems. In: USENIX Annual Technical Conference,
  General Track. pp. 105--120 (2005)

\bibitem{hpnssh}
Rapier, C., Bennett, B.: High speed bulk data transfer using the {SSH}
  protocol. In: Proceedings of the 15th ACM Mardi Gras conference: From
  lightweight mash-ups to lambda grids: Understanding the spectrum of
  distributed computing requirements, applications, tools, infrastructures,
  interoperability, and the incremental adoption of key capabilities. p.~11.
  ACM (2008)

\bibitem{fuse}
Szeredi, M., Rauth, N.: Fuse - filesystems in userspace,
  \url{https://github.com/libfuse/libfuse}

\bibitem{niagara}
Tremblay, M., Chaudhry, S.: A third-generation 65nm 16-core 32-thread plus
  32-scout-thread {CMT SPARC} processor. In: 2008 IEEE International
  Solid-State Circuits Conference-Digest of Technical Papers (2008)

\bibitem{transfs2}
Tweedie, S.C.: Journaling the {Linux} ext2fs filesystem. In: The Fourth Annual
  Linux Expo (1998)

\bibitem{pcachefs}
Tyers, J., Penninckx, P.: {pCacheFS} - persistent-caching {FUSE} filesystem,
  \url{https://github.com/ibizaman/pcachefs}

\bibitem{transfs3}
Wright, C.P., Spillane, R., Sivathanu, G., Zadok, E.: Extending {ACID}
  semantics to the file system. ACM Transactions on Storage (TOS)  3(2), ~4
  (2007)

\end{thebibliography}

\end{document}